\documentclass[%
 reprint,
 amsmath,amssymb,
 aps,
]{revtex4-2}

\usepackage{graphicx}
\usepackage{dcolumn}
\usepackage{bm}
\usepackage{subfiles}

\begin{document}

\preprint{}

\title{Quantum Simulation of the Radical Pair Dynamics of the Avian Compass}

\author{Yiteng Zhang}
 \altaffiliation{Independent Researcher}
\author{Zixuan Hu}
 \altaffiliation{Department of Chemistry, Department of Physics, and Purdue Quantum Science and Engineering Institute}
\author{Yuchen Wang}
 \altaffiliation{Department of Chemistry, Department of Physics, and Purdue Quantum Science and Engineering Institute}
\author{Sabre Kais}
 \email{kais@purdue.edu}
\affiliation{Department of Chemistry, Department of Physics, and Purdue Quantum Science and Engineering Institute}%


\begin{abstract}
The simulation of open quantum dynamics on quantum circuits has attracted wide interests recently with a variety of quantum algorithms developed and demonstrated. Among these, one particular design of a unitary-dilation-based quantum algorithm is capable of simulating general and complex physical systems. In this paper, we apply this quantum algorithm to simulating the dynamics of the radical pair mechanism in the avian compass. This application is demonstrated on the IBM QASM quantum simulator. This work is the first application of any quantum algorithm to simulating the radical pair mechanism in the avian compass, which not only demonstrates the generality of the quantum algorithm, but also opens new opportunities for studying the avian compass with quantum computing devices.
\end{abstract}

\maketitle


\section{\label{sec:level1}Introduction}
An open quantum system is a quantum system that interacts with an external environment or bath. The interaction between the system and the bath is often too complex to be simulated exactly, and thus requires approximations to average out the effects of the bath -- this results in the non-unitary dynamics of open quantum systems. Simulating the dynamics of quantum systems has been a main focus of quantum computing research~\cite{kais2014introduction,cao2019quantum,bauer2020quantum,head2020quantum,huang2020predicting,childs2018toward}, yet relatively few quantum algorithms have been developed for simulating the dynamics of open quantum systems~\cite{wang2011quantum,wang2013solovay,wei2016duality,di2015quantum,sweke2015universal,garcia2020ibm,childs2016efficient,kliesch2011dissipative}. To this end, we have developed and demonstrated a general quantum algorithm for open quantum dynamics \cite{hu1, hu2,wang2022simulation} that is capable of simulating general and complex physical systems. The quantum algorithm leverages the Sz.-Nagy unitary dilation approach to convert non-unitary time evolution operators into corresponding unitary operators, which can then be implemented on a quantum circuit. This quantum algorithm has been applied to a variety of physical systems, including the amplitude damping channel described by the Kraus representation~\cite{hu1}, the Jaynes-Cummings model described by the Kraus representation~\cite{head2021capturing}, the Fenna-Matthews-Olson (FMO) complex described by the Lindblad master equation~\cite{hu2}, and the spin-boson model described by the generalized quantum master equation (GQME)~\cite{wang2022simulation}.

In this work, we apply the general quantum algorithm to simulating the radical pair mechanism in the avian compass and further demonstrate its generality. The radical pair mechanism (RPM) is a theory proposed to explain the magneto-reception and navigation abilities of certain bird species \cite{yt1, yt2, yt3}. Many animals possess extraordinary abilities to sense the direction by perceiving the geomagnetic field. This is probably the result of natural selection over a very long time of evolution, since the ability to sense the direction is crucial for certain animals to find their habitats, such as migratory birds that change habitats from season to season.

In brief, the RPM involves two spatially separated electrons, which are correlated with each other and affected by the external weak magnetic field and internal nuclear spin couplings. The basic scheme of the RPM includes three stages: 1) the photons with certain energies activate a certain type of molecules located in the bird's eyes, enabling an electron transfer reaction and generating a radical pair in the singlet state; 2) the state of the radical pair converts between the singlet state and the triplet state under the influence of the external magnetic field (the geomagnetic field) and the internal magnetic field (the hyperfine coupling effects); 3) the radical pairs in different states will generate different chemical products which can induce a detectable signal for birds to recognize the direction \cite{Hore1}. The RPM is a promising hypothesis that can explain the three unusual properties of the avian compass: 1. the inclination compass: the functional mode of the avian magnetic compass is based on the inclination of the field lines instead of their polarity \cite{inclination1, inclination2, inclination3, inclination4}; 2. the light dependence: light with an energy above a certain threshold is needed for the RPM to work \cite{light1, light2, light3, light4, light5, light6, light7}; 3. the narrow range of responsive magnetic field intensities: both higher and lower magnetic fields will disable birds' ability of navigation \cite{range1}.

To understand the RP mechanism, E. M. Gauger $et.\ al.$ proposed a way to model the dynamics of the RPM system with a Lindblad master equation by adding two ``shelving states" for the singlet yield and triple yield \cite{Gauger1}. In the following, we treat the same Lindblad formulation of the RPM dynamics with our general quantum algorithm for open quantum dynamics and simulate the RPM dynamics on the IBM QASM quantum simulator. To our best knowledge, this is the first ever demonstration of any quantum algorithm applied to simulating the RPM dynamics. This work not only shows the generality of the quantum algorithm, but also opens new potential ways to study the avian compass with quantum computing devices. 

\section{\label{sec:level2}Methodology}
\subsection{The General Quantum Algorithm for Open Quantum Dynamics}
\label{subsec:The General Quantum Algorithm}
The general quantum algorithm for open quantum dynamics has different versions that can evolve the Kraus representation~\cite{hu1}, the Lindblad master equation~\cite{hu2}, and the generalized quantum master equation~\cite{wang2022simulation}. In this work we use the version for the Lindblad master equation~\cite{hu2}.

We first review how the generalized quantum
algorithm converts the non-unitary matrix that encodes the open quantum dynamics into a unitary evaluation based on the Sz.-Nagy unitary
dilation procedure~\cite{hu1,hu2}. We assume the initial density matrix that describes the physical system is composed of a set of unique pure quantum states $|\phi_i\rangle$ that are weighted by their corresponding probabilities $p_i$:
$$
    \rho = \sum_i p_i |\phi_i\rangle\langle \phi_i|
$$
we want to simulate the time evolution of $\rho(t)$ given the initial $\rho$ and the Kraus operators $M_k$’s. This task can be achieved by preparing each input state $|\phi_i\rangle$ in a vector form $v_i$ in a given basis and then building a quantum circuit that generates the quantum state:
\begin{equation}
    |\phi_{ik}(t)\rangle=M_k v_i\xrightarrow{\text{unitary dilation}} U_{M_k}(v_i^T,0,\cdots,0)^T.
\end{equation}
The $U_{M_k}$ is generated via the 1-dilation of $M_k$:
\begin{equation}
\label{eq:1-dialtion}
    U_{M_k}=\begin{pmatrix}
        M_k&D_{M_k^{\dagger}}\\
        D_{M_k}&-M_k^{\dagger}
    \end{pmatrix},
\end{equation}
where $D_{M_k}=\sqrt{I-M_k^{\dagger}M_k},\;D_{M_k^{\dagger}}=\sqrt{I-M_kM_k^{\dagger}}$~\cite{levy2014dilation}.
After obtaining each $|\phi_{ik}(t)\rangle$, we can calculate the population of each basis state in the current basis from the diagonal vector:
\begin{equation}
    \mathrm{diag}(\rho(t))=\sum_{ik} p_i\cdot \mathrm{diag}(|\phi_{ik}(t)\rangle\langle\phi_{ik}(t)|),
    \label{eq: rho_diag}
\end{equation}
where $\mathrm{diag}(|\phi_{ik}(t)\rangle\langle\phi_{ik}(t)|)$ can be efficiently obtained
by applying projection measurements on the first half subspace of $U_{M_k}(v_i^T,0,\cdots,0)^T$.

For the dynamics of an open quantum system, the time evolution of the density matrix can be represented as:
\begin{equation}
\rho(s + \delta s)=\sum_{k}M_{sk}\rho(s)M_{sk}^{\dagger}
\label{eq1}
\end{equation}
where $\rho(s)$ is the density matrix at time step $s$, and $\delta s$ is considered as the discrete time step, during which the Kraus operators $M_{sk}$ are assumed to be constant. The formula eq.(\ref{eq1}) can be used iteratively until reaching the time of interest. Explicitly, the dynamics of the density matrix is described as:
\begin{flalign}
\rho(1) &= \rho(1\delta t) = \sum_{k}M_{0k}\rho(0)M_{0k}^{\dagger} \label{eq:rho1}
\\
\rho(2) &= \rho(2\delta t) = \sum_{k}M_{1k}\rho(1)M_{1k}^{\dagger} &&\label{eq:rho2}\\\nonumber 
&= \sum_{k}\sum_{j}M_{1k}M_{0j}\rho(0)M_{0j}^{\dagger}M_{1k}^{\dagger}&&
\\
\rho(3) &= \rho(3\delta t) = \sum_{k}M_{2k}\rho(2)M_{2k}^{\dagger} &&\label{eq:rho3}\\\nonumber
&= \sum_{k}\sum_{j}\sum_{i}M_{2k}M_{1j}M_{0i}\rho(0)M_{0i}^{\dagger}M_{1j}^{\dagger}M_{2k}^{\dagger}
\\
\dots \nonumber
\end{flalign}
Here without losing any generality, the Kraus operators $M_{sk}$ are indexed by the time step $s$, which allows each time step to have a different set of Kraus operators. However, as will be discussed in Section \ref{subsec:Calculation and Simulation}, the RPM dynamical model used in this work is a Markovian process described by the Lindblad master equation, therefore all the time steps have the same set of Kraus operators $M_{k}$ where the time step index $s$ has been removed.

\subsection{The Radical Pair Mechanism Theory and Dynamics}
\label{subsec:The Radical Pair}
\begin{figure}
    \centering
    \includegraphics[width=0.45\textwidth]{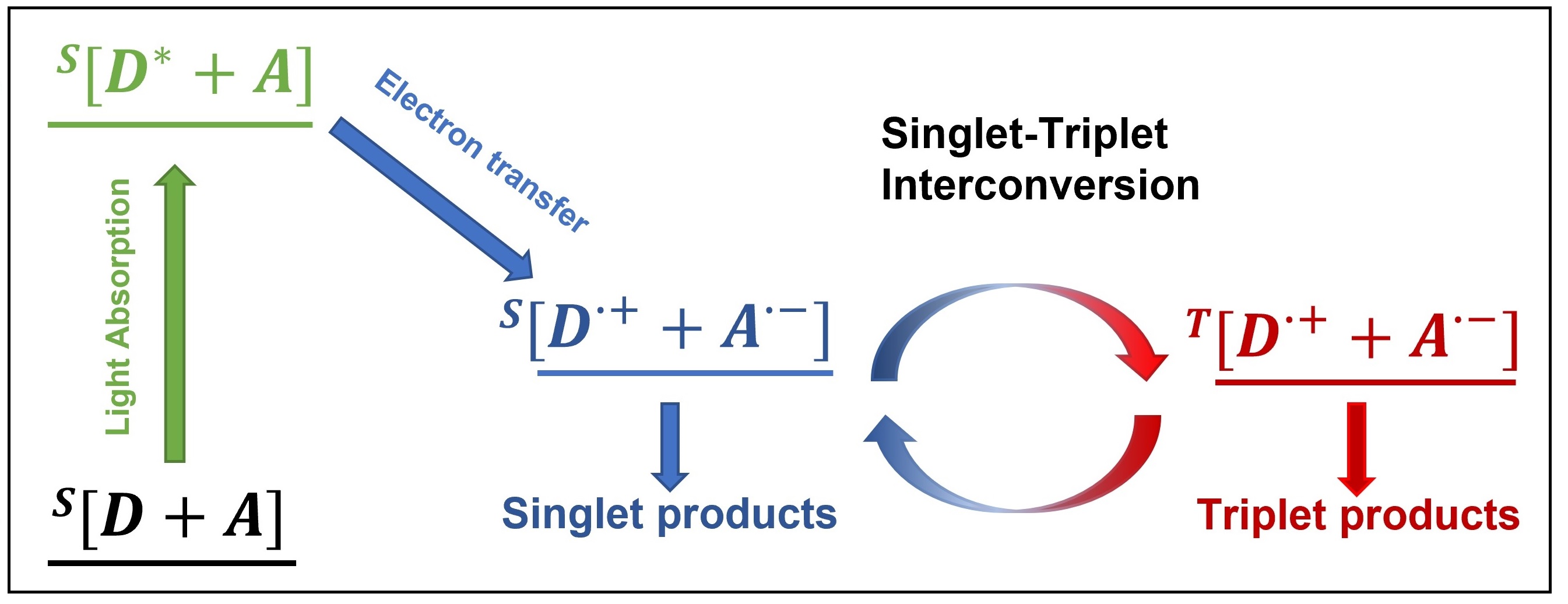}
    \caption{The basic scheme of the radical pair mechanism. After absorbing light, donor (D) and acceptor (A) molecules form radical pairs in its singlet state. Then, under the influence of the magnetic fields, the states of the radical pair interconverse between the singlet states and triplet states. Finally, the singlet and triplet radical pairs end up with different products.}
    \label{fig:rpmscheme}
\end{figure}
The basic scheme of the RPM is shown in Fig. \ref{fig:rpmscheme}. To simplify the fields, we assume that only the electron near the donor interacts with the nucleus, and the electron away from the donor is not affected by the anisotropic hyperfine coupling \cite{Gauger1}. Therefore, the Hamiltonian of the system is
\begin{equation}
    H = \gamma [\hat{I} \cdot \mathbf{A} \cdot \hat{S}_1 + \mathbf{B} \cdot ( \hat{S}_1 + \hat{S}_2 )]
    \label{eq:hamil}
\end{equation}
where $ \mathbf{A} $ is the anisotropic hyperfine tensor coupling the nucleus and one of two spatially separated electrons, and $ \mathbf{A} = diag\{A_x, A_y, A_z\}$ with $A_x = A_y = A_z/2$; $\mathbf{B} = B_0(\cos \varphi \sin \theta, \sin \varphi \sin \theta, \cos \theta)$, and $B_0$ is the magnitude of the geomagnetic field; $\varphi$ is the angle between the x-axis of the radical pair and the external magnetic field; $\theta$ is the angle between the z-axis of the radical pair and the external magnetic field; $\gamma = \frac{1}{2} \mu_0 g$, and $\mu_0$ is the Bohr magneton and $g=2$ is the electron-spin $g$-factor.

To model the dynamics of the system with a quantum master equation formulation, two ``shelving states" were added to the 8-dimensional Hilbert space of the three spins (two electron spins and one nuclear spin) \cite{Gauger1}. We employ operators as shown in Eq.\eqref{eq:projection Op} to represent the spin-selective relaxation into the singlet shelf $|S \rangle$ from the electron singlet state, or the triplet shelf $|T \rangle$ from the electron triplet state. The final populations of $|S \rangle$ and $|T \rangle$ give the singlet and triplet yields. 

With the usual definition of singlet $|s\rangle$ and triplet states $|t_i\rangle$ in the electronic subspace, while $\mid\uparrow\rangle$ and $\mid\downarrow\rangle$ describing the states of the nuclear spin, we define the following decay operators:
\begin{equation}
\begin{aligned}
  P_1 = P_{S, \uparrow} = |S\rangle\langle s,\uparrow|, P_2 = P_{T_0, \uparrow} = |T\rangle\langle t_0,\uparrow| \\
  P_3 = P_{T_+, \uparrow} = |T\rangle\langle t_+,\uparrow|, P_4 = P_{T_-, \uparrow} = |T\rangle\langle t_-,\uparrow|\\
  P_5 = P_{S, \downarrow} = |S\rangle\langle s,\downarrow|, P_6 = P_{T_0, \downarrow} = |T\rangle\langle t_0,\downarrow|\\
  P_7 = P_{T_+, \downarrow} = |T\rangle\langle t_+,\downarrow|, P_8 = P_{T_-, \downarrow} = |T\rangle\langle t_-,\downarrow|
  \label{eq:projection Op}
\end{aligned}
\end{equation}

\begin{table*}
\begin{tabular}{ |p{1.5cm}||p{10cm}||p{3.5cm} | }
 \hline
 \multicolumn{3}{|c|}{Parameter Details} \\
 \hline
 Symbol & Description & Values\\
 \hline
 $A_x$ & Anisotropic hyperfine tensor & $1 \times 10^{-4} T$\\
 \hline
 $B_0$ & Magnitude of the geomagnetic field& $5 \times 10^{-5} T$ \\
 \hline
 $\gamma$ & Half of the product of the Bohr magneton and electron-spin $g$-factor & $9.27 \times 10^{-24} J/T$ \\
 \hline
 $\hbar$ & Reduced Planck constant & $1.05457 \times 10^{-32} J\cdot s$ \\
 \hline
 $\varphi$ & Angle between $x$-axis of the radical pair and the magnetic field& $0$ \\
 \hline
 $k_d$ & Decay Rate of the singlet and triplet states &$1 \times 10^4 s^{-1}$ \\
 \hline
\end{tabular}
\caption{\label{tb:param} Parameter values used in the calculation.}
\end{table*}

This gives a standard Lindblad master equation:
\begin{equation}
    \dot{\rho} = -\frac{i}{\hbar}[H, \rho]+k_d\sum_{i=1}^8 [P_i\rho P_i^{\dagger}-\frac{1}{2}(P_i^{\dagger}P_i\rho + \rho P_i^{\dagger}P_i)],
    \label{eq:me}
\end{equation}
where $k_d$ is the decay rate of the singlet and triplet states. Note the decay rate $k_d$ is independent of the radical pair states, so we have assigned the same decay rate $k_d$ to all eight projectors.

\subsection{Calculation and Simulation}
\label{subsec:Calculation and Simulation}
Now to apply the general quantum algorithm to the RPM dynamics, we first consider the non-unitary part on the right side of Eq. (\ref{eq:me}) which can be rewritten as:
\begin{equation}
    \frac{\delta \rho(t)}{\delta t} = \mathcal{L}(\rho) = k_d\sum_{i=1}^8 [P_i\rho P_i^{\dagger}-\frac{1}{2}\{P_i^{\dagger}P_i, \rho\}] 
    \label{eq:lme}
\end{equation}
Given a very small $\delta t$, Eq. (\ref{eq:lme}) becomes:
\begin{flalign}\label{eq:rhot}
    \rho(t+\delta t) - \rho(t) = &k_d\delta t\sum_{i=1}^8[P_i\rho(t) P_i^{\dagger} - \frac{1}{2}\{P_i^{\dagger}P_i, \rho(t)\}] \\\nonumber
    &+ \mathcal{O}(\delta t^2).
\end{flalign}

\begin{figure}
    \centering
    \includegraphics[width=0.45\textwidth]{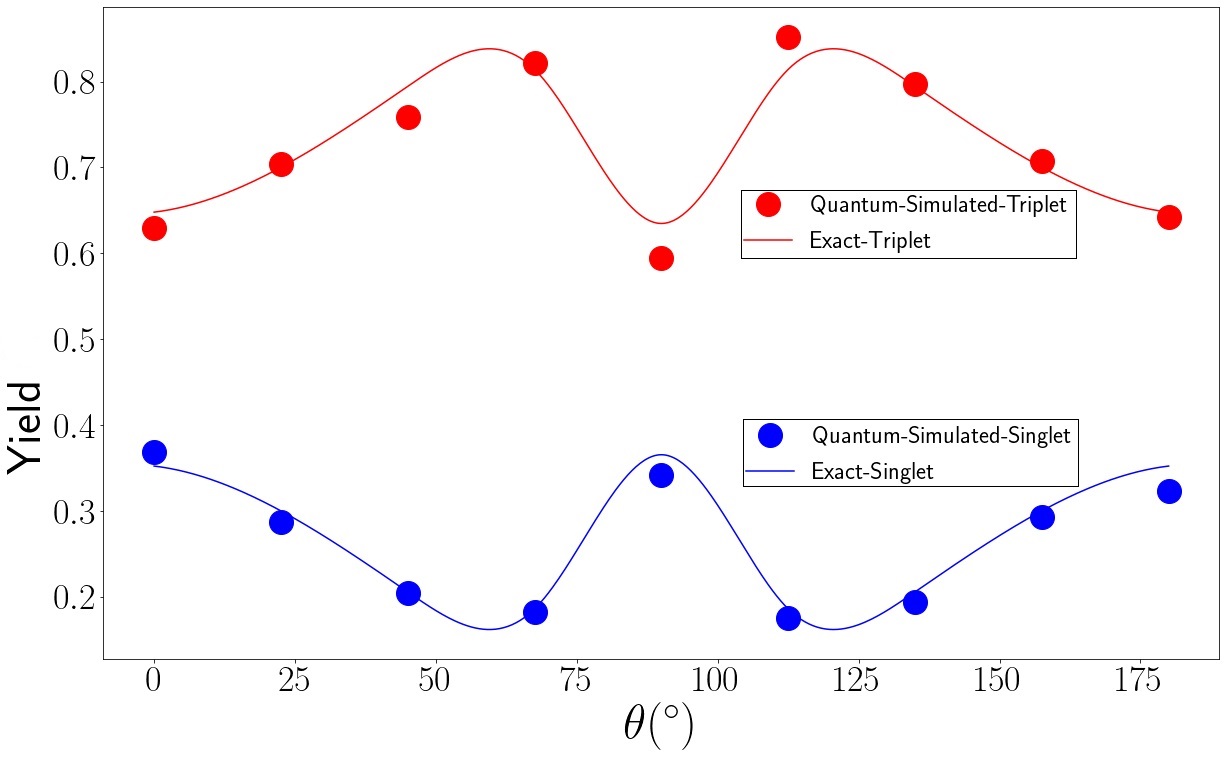}
    \caption{The comparison of the singlet(in blue) and triplet(in red) yields between the results obtained from exact calculation and the quantum simulation of the RPM dynamics. The exact curves are generated from the cubic interpolation of the exact calculation of the yields at each data points. The dots represent the results  simulated by the general quantum algorithm as implemented on the IBM QASM simulator. The parameters used are shown in Table \ref{tb:param}. The yields are calculated around $7.5 \times 10^{-4} s$ after the system has already reached the steady-state. The y-axis is the final singlet/triplet yields -- i.e. the populations of singlet/triplet shelf state; and the x-axis is the angle between z-axis of the radical pair and the magnetic field.}
    \label{fig:qcres}
\end{figure}

\begin{figure}
    \centering
    \includegraphics[width=0.45\textwidth]{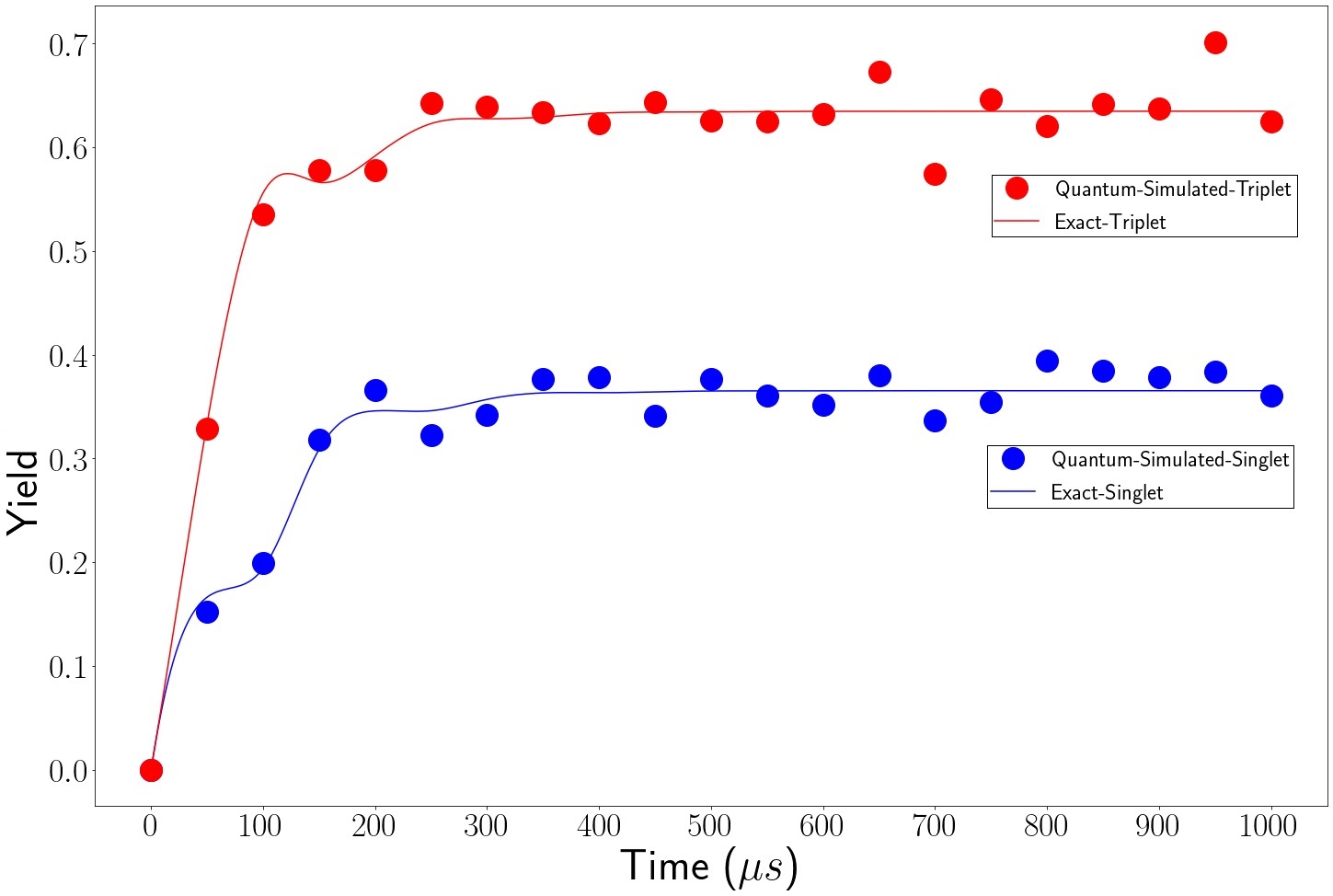}
    \caption{The comparison of the dynamics of the singlet(in blue) and triplet(in red) yields between the exact and quantum simulated results. The exact curves are generated from the cubic interpolation of the exact calculation of the yields at each data points. The dots represent the results  simulated by the general quantum algorithm as implemented on the IBM QASM simulator. After about $2 \times 10^{-4} s$, both yields reach steady-state. The y-axis is the final singlet/triplet yields -- i.e. the populations of singlet/triplet shelf state; and the x-axis is the time.}
    \label{fig:qcdynamics}
\end{figure}

Now assuming $M_0=\sqrt{\mathbf{I} - \frac{1}{2}k_d\delta t\sum_{k=1}^8P_k^{\dagger}P_k}$ and $M_k = \sqrt{k_d\delta t}P_k$ for $k>0$, and ignoring the second order of $\delta t$ as $\delta t \rightarrow 0$, Eq. (\eqref{eq:rhot}) can be rewritten as:
\begin{flalign}
    \rho(t + \delta t) = M_0\rho(t) M_0^{\dagger} + \sum_{k=1}^8 M_k \rho(t) M_k^{\dagger}
    \label{eq:kraus}
\end{flalign}
Eq.(\ref{eq:kraus}) is in the same form of Eq.(\ref{eq1}), thus knowing the initial state $\rho(0)$ we can evolve the density matrix to a certain time with the iterative procedure described in Eq.\eqref{eq:rho1} to Eq.\eqref{eq:rho3}.

In the meantime,
\begin{flalign}
    \sum_{k=0}^8 M_k^{\dagger} M_k &= M_0^{\dagger}M_0 + \sum_{k=1}^8 M_k^{\dagger}M_k \\ \nonumber
    &= \mathbf{I} - k_d\delta t \sum_{k=1}^8 P_k^{\dagger}P_k + + \mathcal{O}(\delta t^2) + k_d \delta t\sum_{k=1}^8 P_k^{\dagger}P_k \\ \nonumber
    &= \mathbf{I} + \mathcal{O}(\delta t^2)
    \label{eq:msum}
\end{flalign}

When $\delta t \rightarrow 0$, according to the above equation, we have:
\begin{equation}
    M_0 = \sqrt{\mathbf{I} - k_d \delta t\sum_{k=1}^8 P_k^{\dagger}P_k}
    \label{eq:m0}
\end{equation}
With Eq. (\ref{eq:m0}), we can formulate the Kraus operators to satisfy the condition $\sum_k M_k^{\dagger}M_k = \mathbf{I} $. Thus, we have defined all the 9 Kraus operators $M_k$ required to describe the RPM dynamics.

There is one additional term containing the Hamiltonian, $-\frac{i}{\hbar}[H, \rho]$, in Eq. (\ref{eq:me}) as compared with Eq. (\ref{eq:lme}). This ``oscillating part" of the dynamics is unitary and thus can be easily realized by multiplying each Kraus operator by a unitary matrix obtained through the diagonalization of the Hamiltonian \cite{hu2}.

With the parameters in Table \ref{tb:param}, we simulated the RPM dynamics by iteratively applying Eq. (\ref{eq:kraus}) on the IBM QASM quantum simulator and then used the output results to calculate the singlet and triplet yields. Also, we assume the initial state of the two electron spins is $\frac{1}{\sqrt{2}}(\mid\uparrow\downarrow\rangle - \mid\downarrow\uparrow\rangle)$, and the initial state of the nuclear spin is $\frac{1}{\sqrt{2}}(\mid\uparrow\rangle - \mid\downarrow\rangle)$. In our simulation, the time interval $\delta t$ is set as $0.5/k_d=5\times10^{-5}s$. We then apply the procedure shown in Eq.(\ref{eq:rho1}), Eq.(\ref{eq:rho2}) and Eq.(\ref{eq:rho3}). As mentioned above, the populations of $\mid S \rangle$ and $\mid T \rangle$ is the singlet and triplet yields respectively, after the system reaches the steady state. The populations are calculated by the procedure explained in Eq. (\ref{eq: rho_diag}), where the diagonal elements of the density matrix are obtained by projection measurements into the computational subspace.

The results are compared with those obtained from classical methods in Fig. \ref{fig:qcres}, where the quantum algorithm results are highly consistent with the classical method results. Fig. \ref{fig:qcdynamics} shows the dynamical evolution of the singlet and triplet yields (when $\theta = \frac{\pi}{2}$) as simulated by the general quantum algorithm on the IBM quantum simulator. After $2 \times 10^{-4} s$, the yields almost reach steady-state, which is consistent with the chosen decay rate of $k=1\times 10^{4} s^{-1}$.

\subsection{Complexity analysis}
\label{subsec:complexity analysis}

One factor that contributes to the complexity of the quantum algorithm is the system's size. For a density matrix of the size $n\times n$, the cost to realize the unitary dilation of a most general $n\times n$ Kraus operator $M_k$ is $\mathcal{O}(n^2)$~\cite{hu1,hu2}. However, in our calculation of the dynamics of the RPM, the Kraus operators each represents a single elementary physical process and thus the $M_k$ matrices are often sparse with few non-zero elements. This means the practical complexity scaling of implementing each $M_k$ matrix on a quantum circuit can be greatly reduced to $\mathcal{O}(\log^2 n)$. Taking into account the total $K$ number of $M_k$ matrices to be simulated on the quantum circuit, the total complexity scaling is $\mathcal{O}(K\log^2 n)$ for our given system. It is worth noting that the $K$ is determined on a case-by-case basis by the dynamical model and different $M_k$ matrices can be evolved in parallel, therefore the scaling in $K$ is a ``soft" scaling that does not contribute to either the depth or the width of each individual quantum circuit~\cite{hu1,hu2}. Another contributing factor to the quantum algorithm's complexity is the number of time steps. In the most general case, as can be seen from Eq.\eqref{eq:rho1} to Eq.\eqref{eq:rho3}, taking $s$ steps requires $K^s$ matrices to be evolved, which is an exponential scaling in the number of time steps. However, fortunately the actual number of matrix terms to be simulated can be greatly reduced once again due to the sparsity of the $M_k$ matrices. As mentioned above, the Kraus operators represent elementary physical processes and thus the $M_k$ matrices are often sparse with very few non-zero elements: this means that most matrix product terms in e.g. Eq.\eqref{eq:rho3} are zero matrices or matrices with negligible norms. The actual number of matrix products we need to evaluate is determined on a case-by-case basis. In the current simulation, in theory the total number of terms in $n$-th iteration will be $9^n$ with nine Kraus operators $\{M_k|k=0, 1, 2,...,8\}$. However, since the product of each pair of the the decay operators $\{P_k|k=1, 2, 3,...,8\}$ is $0$, the product of each pair of the Kraus operator $\{M_k|k=1, 2, 3,...,8\}$ is 0. Therefore, there will be only 8 more terms when adding one more iteration. Thus, there will be $8 \times n + 1$ terms in $n$-th iteration, reducing the terms significantly. More simulation details are in the Supplementary Information.

\section{\label{sec:level3}Conclusion}
Based on the calculations and results, we can conclude that the general quantum algorithm based on the Sz.-Nagy dilation can accurately simulate the RPM dynamics described by the Lindblad master equation. As discussed above, the RPM is an important theory that can explain the magneto-reception process of the avian compass. 
The RPM also acts as an ideal model to help explaining the isotope effects in xenon anaesthesia and lithium treatment of
hyperactivity, magnetic field effects on the circadian clock, as well as hypomagnetic
field effects on neurogenesis and microtubule assembly~\cite{kominis2015radical,zadeh2022magnetic}. Our simulation of the RPM model with the quantum algorithm not only demonstrates the generality of the algorithm but also helps bridging the gap between applying tools of quantum-information science to the investigation of new areas of quantum biology.


 \begin{acknowledgments}
We acknowledge the financial support of  the National Science Foundation under award number 2124511, CCI Phase I: NSF Center for Quantum Dynamics on Modular Quantum Devices (CQD-MQD). We also acknowledge the financial support from the U.S. Department of Energy under Award No. DE-SC0019215.
We acknowledge the use of IBM Quantum services for this work.
The views expressed are those of the authors, and do not
reflect the official policy or position of IBM or the IBM
Quantum team.
 \end{acknowledgments}


\nocite{*}
\newpage
\bibliography{quantum_algo_rpm}

\providecommand{\noopsort}[1]{}\providecommand{\singleletter}[1]{#1}%
\begin{thebibliography}{38}%
\makeatletter
\providecommand \@ifxundefined [1]{%
 \@ifx{#1\undefined}
}%
\providecommand \@ifnum [1]{%
 \ifnum #1\expandafter \@firstoftwo
 \else \expandafter \@secondoftwo
 \fi
}%
\providecommand \@ifx [1]{%
 \ifx #1\expandafter \@firstoftwo
 \else \expandafter \@secondoftwo
 \fi
}%
\providecommand \natexlab [1]{#1}%
\providecommand \enquote  [1]{``#1''}%
\providecommand \bibnamefont  [1]{#1}%
\providecommand \bibfnamefont [1]{#1}%
\providecommand \citenamefont [1]{#1}%
\providecommand \href@noop [0]{\@secondoftwo}%
\providecommand \href [0]{\begingroup \@sanitize@url \@href}%
\providecommand \@href[1]{\@@startlink{#1}\@@href}%
\providecommand \@@href[1]{\endgroup#1\@@endlink}%
\providecommand \@sanitize@url [0]{\catcode `\\12\catcode `\$12\catcode
  `\&12\catcode `\#12\catcode `\^12\catcode `\_12\catcode `\%12\relax}%
\providecommand \@@startlink[1]{}%
\providecommand \@@endlink[0]{}%
\providecommand \url  [0]{\begingroup\@sanitize@url \@url }%
\providecommand \@url [1]{\endgroup\@href {#1}{\urlprefix }}%
\providecommand \urlprefix  [0]{URL }%
\providecommand \Eprint [0]{\href }%
\providecommand \doibase [0]{https://doi.org/}%
\providecommand \selectlanguage [0]{\@gobble}%
\providecommand \bibinfo  [0]{\@secondoftwo}%
\providecommand \bibfield  [0]{\@secondoftwo}%
\providecommand \translation [1]{[#1]}%
\providecommand \BibitemOpen [0]{}%
\providecommand \bibitemStop [0]{}%
\providecommand \bibitemNoStop [0]{.\EOS\space}%
\providecommand \EOS [0]{\spacefactor3000\relax}%
\providecommand \BibitemShut  [1]{\csname bibitem#1\endcsname}%
\let\auto@bib@innerbib\@empty
\bibitem [{\citenamefont {Kais}(2014)}]{kais2014introduction}%
  \BibitemOpen
  \bibfield  {author} {\bibinfo {author} {\bibfnamefont {S.}~\bibnamefont
  {Kais}},\ }\bibfield  {title} {\bibinfo {title} {Introduction to quantum
  information and computation for chemistry},\ }\href@noop {} {\bibfield
  {journal} {\bibinfo  {journal} {Quantum Information and Computation for
  Chemistry}\ ,\ \bibinfo {pages} {1}} (\bibinfo {year} {2014})}\BibitemShut
  {NoStop}%
\bibitem [{\citenamefont {Cao}\ \emph {et~al.}(2019)\citenamefont {Cao},
  \citenamefont {Romero}, \citenamefont {Olson}, \citenamefont {Degroote},
  \citenamefont {Johnson}, \citenamefont {Kieferov{\'a}}, \citenamefont
  {Kivlichan}, \citenamefont {Menke}, \citenamefont {Peropadre}, \citenamefont
  {Sawaya} \emph {et~al.}}]{cao2019quantum}%
  \BibitemOpen
  \bibfield  {author} {\bibinfo {author} {\bibfnamefont {Y.}~\bibnamefont
  {Cao}}, \bibinfo {author} {\bibfnamefont {J.}~\bibnamefont {Romero}},
  \bibinfo {author} {\bibfnamefont {J.~P.}\ \bibnamefont {Olson}}, \bibinfo
  {author} {\bibfnamefont {M.}~\bibnamefont {Degroote}}, \bibinfo {author}
  {\bibfnamefont {P.~D.}\ \bibnamefont {Johnson}}, \bibinfo {author}
  {\bibfnamefont {M.}~\bibnamefont {Kieferov{\'a}}}, \bibinfo {author}
  {\bibfnamefont {I.~D.}\ \bibnamefont {Kivlichan}}, \bibinfo {author}
  {\bibfnamefont {T.}~\bibnamefont {Menke}}, \bibinfo {author} {\bibfnamefont
  {B.}~\bibnamefont {Peropadre}}, \bibinfo {author} {\bibfnamefont {N.~P.}\
  \bibnamefont {Sawaya}}, \emph {et~al.},\ }\bibfield  {title} {\bibinfo
  {title} {Quantum chemistry in the age of quantum computing},\ }\href@noop {}
  {\bibfield  {journal} {\bibinfo  {journal} {Chemical reviews}\ }\textbf
  {\bibinfo {volume} {119}},\ \bibinfo {pages} {10856} (\bibinfo {year}
  {2019})}\BibitemShut {NoStop}%
\bibitem [{\citenamefont {Bauer}\ \emph {et~al.}(2020)\citenamefont {Bauer},
  \citenamefont {Bravyi}, \citenamefont {Motta},\ and\ \citenamefont
  {Chan}}]{bauer2020quantum}%
  \BibitemOpen
  \bibfield  {author} {\bibinfo {author} {\bibfnamefont {B.}~\bibnamefont
  {Bauer}}, \bibinfo {author} {\bibfnamefont {S.}~\bibnamefont {Bravyi}},
  \bibinfo {author} {\bibfnamefont {M.}~\bibnamefont {Motta}},\ and\ \bibinfo
  {author} {\bibfnamefont {G.~K.-L.}\ \bibnamefont {Chan}},\ }\bibfield
  {title} {\bibinfo {title} {Quantum algorithms for quantum chemistry and
  quantum materials science},\ }\href@noop {} {\bibfield  {journal} {\bibinfo
  {journal} {Chemical Reviews}\ }\textbf {\bibinfo {volume} {120}},\ \bibinfo
  {pages} {12685} (\bibinfo {year} {2020})}\BibitemShut {NoStop}%
\bibitem [{\citenamefont {Head-Marsden}\ \emph {et~al.}(2020)\citenamefont
  {Head-Marsden}, \citenamefont {Flick}, \citenamefont {Ciccarino},\ and\
  \citenamefont {Narang}}]{head2020quantum}%
  \BibitemOpen
  \bibfield  {author} {\bibinfo {author} {\bibfnamefont {K.}~\bibnamefont
  {Head-Marsden}}, \bibinfo {author} {\bibfnamefont {J.}~\bibnamefont {Flick}},
  \bibinfo {author} {\bibfnamefont {C.~J.}\ \bibnamefont {Ciccarino}},\ and\
  \bibinfo {author} {\bibfnamefont {P.}~\bibnamefont {Narang}},\ }\bibfield
  {title} {\bibinfo {title} {Quantum information and algorithms for correlated
  quantum matter},\ }\href@noop {} {\bibfield  {journal} {\bibinfo  {journal}
  {Chemical Reviews}\ }\textbf {\bibinfo {volume} {121}},\ \bibinfo {pages}
  {3061} (\bibinfo {year} {2020})}\BibitemShut {NoStop}%
\bibitem [{\citenamefont {Huang}\ \emph {et~al.}(2020)\citenamefont {Huang},
  \citenamefont {Kueng},\ and\ \citenamefont {Preskill}}]{huang2020predicting}%
  \BibitemOpen
  \bibfield  {author} {\bibinfo {author} {\bibfnamefont {H.-Y.}\ \bibnamefont
  {Huang}}, \bibinfo {author} {\bibfnamefont {R.}~\bibnamefont {Kueng}},\ and\
  \bibinfo {author} {\bibfnamefont {J.}~\bibnamefont {Preskill}},\ }\bibfield
  {title} {\bibinfo {title} {Predicting many properties of a quantum system
  from very few measurements},\ }\href@noop {} {\bibfield  {journal} {\bibinfo
  {journal} {Nature Physics}\ }\textbf {\bibinfo {volume} {16}},\ \bibinfo
  {pages} {1050} (\bibinfo {year} {2020})}\BibitemShut {NoStop}%
\bibitem [{\citenamefont {Childs}\ \emph {et~al.}(2018)\citenamefont {Childs},
  \citenamefont {Maslov}, \citenamefont {Nam}, \citenamefont {Ross},\ and\
  \citenamefont {Su}}]{childs2018toward}%
  \BibitemOpen
  \bibfield  {author} {\bibinfo {author} {\bibfnamefont {A.~M.}\ \bibnamefont
  {Childs}}, \bibinfo {author} {\bibfnamefont {D.}~\bibnamefont {Maslov}},
  \bibinfo {author} {\bibfnamefont {Y.}~\bibnamefont {Nam}}, \bibinfo {author}
  {\bibfnamefont {N.~J.}\ \bibnamefont {Ross}},\ and\ \bibinfo {author}
  {\bibfnamefont {Y.}~\bibnamefont {Su}},\ }\bibfield  {title} {\bibinfo
  {title} {Toward the first quantum simulation with quantum speedup},\
  }\href@noop {} {\bibfield  {journal} {\bibinfo  {journal} {Proceedings of the
  National Academy of Sciences}\ }\textbf {\bibinfo {volume} {115}},\ \bibinfo
  {pages} {9456} (\bibinfo {year} {2018})}\BibitemShut {NoStop}%
\bibitem [{\citenamefont {Wang}\ \emph {et~al.}(2011)\citenamefont {Wang},
  \citenamefont {Ashhab},\ and\ \citenamefont {Nori}}]{wang2011quantum}%
  \BibitemOpen
  \bibfield  {author} {\bibinfo {author} {\bibfnamefont {H.}~\bibnamefont
  {Wang}}, \bibinfo {author} {\bibfnamefont {S.}~\bibnamefont {Ashhab}},\ and\
  \bibinfo {author} {\bibfnamefont {F.}~\bibnamefont {Nori}},\ }\bibfield
  {title} {\bibinfo {title} {Quantum algorithm for simulating the dynamics of
  an open quantum system},\ }\href@noop {} {\bibfield  {journal} {\bibinfo
  {journal} {Physical Review A}\ }\textbf {\bibinfo {volume} {83}},\ \bibinfo
  {pages} {062317} (\bibinfo {year} {2011})}\BibitemShut {NoStop}%
\bibitem [{\citenamefont {Wang}\ \emph {et~al.}(2013)\citenamefont {Wang},
  \citenamefont {Berry}, \citenamefont {De~Oliveira},\ and\ \citenamefont
  {Sanders}}]{wang2013solovay}%
  \BibitemOpen
  \bibfield  {author} {\bibinfo {author} {\bibfnamefont {D.-S.}\ \bibnamefont
  {Wang}}, \bibinfo {author} {\bibfnamefont {D.~W.}\ \bibnamefont {Berry}},
  \bibinfo {author} {\bibfnamefont {M.~C.}\ \bibnamefont {De~Oliveira}},\ and\
  \bibinfo {author} {\bibfnamefont {B.~C.}\ \bibnamefont {Sanders}},\
  }\bibfield  {title} {\bibinfo {title} {Solovay-kitaev decomposition strategy
  for single-qubit channels},\ }\href@noop {} {\bibfield  {journal} {\bibinfo
  {journal} {Physical review letters}\ }\textbf {\bibinfo {volume} {111}},\
  \bibinfo {pages} {130504} (\bibinfo {year} {2013})}\BibitemShut {NoStop}%
\bibitem [{\citenamefont {Wei}\ \emph {et~al.}(2016)\citenamefont {Wei},
  \citenamefont {Ruan},\ and\ \citenamefont {Long}}]{wei2016duality}%
  \BibitemOpen
  \bibfield  {author} {\bibinfo {author} {\bibfnamefont {S.-J.}\ \bibnamefont
  {Wei}}, \bibinfo {author} {\bibfnamefont {D.}~\bibnamefont {Ruan}},\ and\
  \bibinfo {author} {\bibfnamefont {G.-L.}\ \bibnamefont {Long}},\ }\bibfield
  {title} {\bibinfo {title} {Duality quantum algorithm efficiently simulates
  open quantum systems},\ }\href@noop {} {\bibfield  {journal} {\bibinfo
  {journal} {Scientific Reports}\ }\textbf {\bibinfo {volume} {6}},\ \bibinfo
  {pages} {1} (\bibinfo {year} {2016})}\BibitemShut {NoStop}%
\bibitem [{\citenamefont {Di~Candia}\ \emph {et~al.}(2015)\citenamefont
  {Di~Candia}, \citenamefont {Pedernales}, \citenamefont {Del~Campo},
  \citenamefont {Solano},\ and\ \citenamefont {Casanova}}]{di2015quantum}%
  \BibitemOpen
  \bibfield  {author} {\bibinfo {author} {\bibfnamefont {R.}~\bibnamefont
  {Di~Candia}}, \bibinfo {author} {\bibfnamefont {J.~S.}\ \bibnamefont
  {Pedernales}}, \bibinfo {author} {\bibfnamefont {A.}~\bibnamefont
  {Del~Campo}}, \bibinfo {author} {\bibfnamefont {E.}~\bibnamefont {Solano}},\
  and\ \bibinfo {author} {\bibfnamefont {J.}~\bibnamefont {Casanova}},\
  }\bibfield  {title} {\bibinfo {title} {Quantum simulation of dissipative
  processes without reservoir engineering},\ }\href@noop {} {\bibfield
  {journal} {\bibinfo  {journal} {Scientific reports}\ }\textbf {\bibinfo
  {volume} {5}},\ \bibinfo {pages} {1} (\bibinfo {year} {2015})}\BibitemShut
  {NoStop}%
\bibitem [{\citenamefont {Sweke}\ \emph {et~al.}(2015)\citenamefont {Sweke},
  \citenamefont {Sinayskiy}, \citenamefont {Bernard},\ and\ \citenamefont
  {Petruccione}}]{sweke2015universal}%
  \BibitemOpen
  \bibfield  {author} {\bibinfo {author} {\bibfnamefont {R.}~\bibnamefont
  {Sweke}}, \bibinfo {author} {\bibfnamefont {I.}~\bibnamefont {Sinayskiy}},
  \bibinfo {author} {\bibfnamefont {D.}~\bibnamefont {Bernard}},\ and\ \bibinfo
  {author} {\bibfnamefont {F.}~\bibnamefont {Petruccione}},\ }\bibfield
  {title} {\bibinfo {title} {Universal simulation of markovian open quantum
  systems},\ }\href@noop {} {\bibfield  {journal} {\bibinfo  {journal}
  {Physical Review A}\ }\textbf {\bibinfo {volume} {91}},\ \bibinfo {pages}
  {062308} (\bibinfo {year} {2015})}\BibitemShut {NoStop}%
\bibitem [{\citenamefont {Garc{\'\i}a-P{\'e}rez}\ \emph
  {et~al.}(2020)\citenamefont {Garc{\'\i}a-P{\'e}rez}, \citenamefont {Rossi},\
  and\ \citenamefont {Maniscalco}}]{garcia2020ibm}%
  \BibitemOpen
  \bibfield  {author} {\bibinfo {author} {\bibfnamefont {G.}~\bibnamefont
  {Garc{\'\i}a-P{\'e}rez}}, \bibinfo {author} {\bibfnamefont {M.~A.}\
  \bibnamefont {Rossi}},\ and\ \bibinfo {author} {\bibfnamefont
  {S.}~\bibnamefont {Maniscalco}},\ }\bibfield  {title} {\bibinfo {title} {Ibm
  q experience as a versatile experimental testbed for simulating open quantum
  systems},\ }\href@noop {} {\bibfield  {journal} {\bibinfo  {journal} {npj
  Quantum Information}\ }\textbf {\bibinfo {volume} {6}},\ \bibinfo {pages} {1}
  (\bibinfo {year} {2020})}\BibitemShut {NoStop}%
\bibitem [{\citenamefont {Childs}\ and\ \citenamefont
  {Li}(2016)}]{childs2016efficient}%
  \BibitemOpen
  \bibfield  {author} {\bibinfo {author} {\bibfnamefont {A.~M.}\ \bibnamefont
  {Childs}}\ and\ \bibinfo {author} {\bibfnamefont {T.}~\bibnamefont {Li}},\
  }\bibfield  {title} {\bibinfo {title} {Efficient simulation of sparse
  markovian quantum dynamics},\ }\href@noop {} {\bibfield  {journal} {\bibinfo
  {journal} {arXiv preprint arXiv:1611.05543}\ } (\bibinfo {year}
  {2016})}\BibitemShut {NoStop}%
\bibitem [{\citenamefont {Kliesch}\ \emph {et~al.}(2011)\citenamefont
  {Kliesch}, \citenamefont {Barthel}, \citenamefont {Gogolin}, \citenamefont
  {Kastoryano},\ and\ \citenamefont {Eisert}}]{kliesch2011dissipative}%
  \BibitemOpen
  \bibfield  {author} {\bibinfo {author} {\bibfnamefont {M.}~\bibnamefont
  {Kliesch}}, \bibinfo {author} {\bibfnamefont {T.}~\bibnamefont {Barthel}},
  \bibinfo {author} {\bibfnamefont {C.}~\bibnamefont {Gogolin}}, \bibinfo
  {author} {\bibfnamefont {M.}~\bibnamefont {Kastoryano}},\ and\ \bibinfo
  {author} {\bibfnamefont {J.}~\bibnamefont {Eisert}},\ }\bibfield  {title}
  {\bibinfo {title} {Dissipative quantum church-turing theorem},\ }\href@noop
  {} {\bibfield  {journal} {\bibinfo  {journal} {Physical review letters}\
  }\textbf {\bibinfo {volume} {107}},\ \bibinfo {pages} {120501} (\bibinfo
  {year} {2011})}\BibitemShut {NoStop}%
\bibitem [{\citenamefont {Hu}\ \emph {et~al.}(2020)\citenamefont {Hu},
  \citenamefont {Xia},\ and\ \citenamefont {Kais}}]{hu1}%
  \BibitemOpen
  \bibfield  {author} {\bibinfo {author} {\bibfnamefont {Z.}~\bibnamefont
  {Hu}}, \bibinfo {author} {\bibfnamefont {R.}~\bibnamefont {Xia}},\ and\
  \bibinfo {author} {\bibfnamefont {S.}~\bibnamefont {Kais}},\ }\bibfield
  {title} {\bibinfo {title} {A quantum algorithm for evolving open quantum
  dynamics on quantum computing devices},\ }\href@noop {} {\bibfield  {journal}
  {\bibinfo  {journal} {Scientific Reports}\ }\textbf {\bibinfo {volume}
  {10}},\ \bibinfo {pages} {3301} (\bibinfo {year} {2020})}\BibitemShut
  {NoStop}%
\bibitem [{\citenamefont {Hu}\ \emph {et~al.}(2022)\citenamefont {Hu},
  \citenamefont {Head-Marsden}, \citenamefont {Mazziotti}, \citenamefont
  {Narang},\ and\ \citenamefont {Kais}}]{hu2}%
  \BibitemOpen
  \bibfield  {author} {\bibinfo {author} {\bibfnamefont {Z.}~\bibnamefont
  {Hu}}, \bibinfo {author} {\bibfnamefont {K.}~\bibnamefont {Head-Marsden}},
  \bibinfo {author} {\bibfnamefont {D.~A.}\ \bibnamefont {Mazziotti}}, \bibinfo
  {author} {\bibfnamefont {P.}~\bibnamefont {Narang}},\ and\ \bibinfo {author}
  {\bibfnamefont {S.}~\bibnamefont {Kais}},\ }\bibfield  {title} {\bibinfo
  {title} {A general quantum algorithm for open quantum dynamics demonstrated
  with the fenna-matthews-olson complex},\ }\href@noop {} {\bibfield  {journal}
  {\bibinfo  {journal} {Quantum}\ }\textbf {\bibinfo {volume} {6}},\ \bibinfo
  {pages} {726} (\bibinfo {year} {2022})}\BibitemShut {NoStop}%
\bibitem [{\citenamefont {Wang}\ \emph {et~al.}(2022)\citenamefont {Wang},
  \citenamefont {Mulvihill}, \citenamefont {Hu}, \citenamefont {Lyu},
  \citenamefont {Shivpuje}, \citenamefont {Liu}, \citenamefont {Soley},
  \citenamefont {Geva}, \citenamefont {Batista},\ and\ \citenamefont
  {Kais}}]{wang2022simulation}%
  \BibitemOpen
  \bibfield  {author} {\bibinfo {author} {\bibfnamefont {Y.}~\bibnamefont
  {Wang}}, \bibinfo {author} {\bibfnamefont {E.}~\bibnamefont {Mulvihill}},
  \bibinfo {author} {\bibfnamefont {Z.}~\bibnamefont {Hu}}, \bibinfo {author}
  {\bibfnamefont {N.}~\bibnamefont {Lyu}}, \bibinfo {author} {\bibfnamefont
  {S.}~\bibnamefont {Shivpuje}}, \bibinfo {author} {\bibfnamefont
  {Y.}~\bibnamefont {Liu}}, \bibinfo {author} {\bibfnamefont {M.~B.}\
  \bibnamefont {Soley}}, \bibinfo {author} {\bibfnamefont {E.}~\bibnamefont
  {Geva}}, \bibinfo {author} {\bibfnamefont {V.~S.}\ \bibnamefont {Batista}},\
  and\ \bibinfo {author} {\bibfnamefont {S.}~\bibnamefont {Kais}},\ }\bibfield
  {title} {\bibinfo {title} {Simulation of open quantum system dynamics based
  on the generalized quantum master equation on quantum computing devices},\
  }\href@noop {} {\bibfield  {journal} {\bibinfo  {journal} {arXiv preprint
  arXiv:2209.04956}\ } (\bibinfo {year} {2022})}\BibitemShut {NoStop}%
\bibitem [{\citenamefont {Head-Marsden}\ \emph {et~al.}(2021)\citenamefont
  {Head-Marsden}, \citenamefont {Krastanov}, \citenamefont {Mazziotti},\ and\
  \citenamefont {Narang}}]{head2021capturing}%
  \BibitemOpen
  \bibfield  {author} {\bibinfo {author} {\bibfnamefont {K.}~\bibnamefont
  {Head-Marsden}}, \bibinfo {author} {\bibfnamefont {S.}~\bibnamefont
  {Krastanov}}, \bibinfo {author} {\bibfnamefont {D.~A.}\ \bibnamefont
  {Mazziotti}},\ and\ \bibinfo {author} {\bibfnamefont {P.}~\bibnamefont
  {Narang}},\ }\bibfield  {title} {\bibinfo {title} {Capturing non-markovian
  dynamics on near-term quantum computers},\ }\href@noop {} {\bibfield
  {journal} {\bibinfo  {journal} {Physical Review Research}\ }\textbf {\bibinfo
  {volume} {3}},\ \bibinfo {pages} {013182} (\bibinfo {year}
  {2021})}\BibitemShut {NoStop}%
\bibitem [{\citenamefont {Pauls}\ \emph {et~al.}(2013)\citenamefont {Pauls},
  \citenamefont {Zhang}, \citenamefont {Berman},\ and\ \citenamefont
  {Kais}}]{yt1}%
  \BibitemOpen
  \bibfield  {author} {\bibinfo {author} {\bibfnamefont {J.~A.}\ \bibnamefont
  {Pauls}}, \bibinfo {author} {\bibfnamefont {Y.}~\bibnamefont {Zhang}},
  \bibinfo {author} {\bibfnamefont {G.~P.}\ \bibnamefont {Berman}},\ and\
  \bibinfo {author} {\bibfnamefont {S.}~\bibnamefont {Kais}},\ }\bibfield
  {title} {\bibinfo {title} {Quantum coherence and entanglement in the avian
  compass},\ }\href@noop {} {\bibfield  {journal} {\bibinfo  {journal}
  {Physical Review E}\ }\textbf {\bibinfo {volume} {87}},\ \bibinfo {pages}
  {062704} (\bibinfo {year} {2013})}\BibitemShut {NoStop}%
\bibitem [{\citenamefont {Zhang}\ \emph
  {et~al.}(2014{\natexlab{a}})\citenamefont {Zhang}, \citenamefont {Berman},\
  and\ \citenamefont {Kais}}]{yt2}%
  \BibitemOpen
  \bibfield  {author} {\bibinfo {author} {\bibfnamefont {Y.}~\bibnamefont
  {Zhang}}, \bibinfo {author} {\bibfnamefont {G.~P.}\ \bibnamefont {Berman}},\
  and\ \bibinfo {author} {\bibfnamefont {S.}~\bibnamefont {Kais}},\ }\bibfield
  {title} {\bibinfo {title} {Sensitivity and entanglement in the avian chemical
  compass},\ }\href@noop {} {\bibfield  {journal} {\bibinfo  {journal}
  {Physical Review E}\ }\textbf {\bibinfo {volume} {90}},\ \bibinfo {pages}
  {042707} (\bibinfo {year} {2014}{\natexlab{a}})}\BibitemShut {NoStop}%
\bibitem [{\citenamefont {Zhang}\ \emph
  {et~al.}(2014{\natexlab{b}})\citenamefont {Zhang}, \citenamefont {Berman},\
  and\ \citenamefont {Kais}}]{yt3}%
  \BibitemOpen
  \bibfield  {author} {\bibinfo {author} {\bibfnamefont {Y.}~\bibnamefont
  {Zhang}}, \bibinfo {author} {\bibfnamefont {G.~P.}\ \bibnamefont {Berman}},\
  and\ \bibinfo {author} {\bibfnamefont {S.}~\bibnamefont {Kais}},\ }\bibfield
  {title} {\bibinfo {title} {The radical pair mechanism and the avian chemical
  compass: Quantum coherence and entanglement},\ }\href@noop {} {\bibfield
  {journal} {\bibinfo  {journal} {International Journal of Quantum Chemistry}\
  }\textbf {\bibinfo {volume} {115}},\ \bibinfo {pages} {1327} (\bibinfo {year}
  {2014}{\natexlab{b}})}\BibitemShut {NoStop}%
\bibitem [{\citenamefont {Dodson}\ \emph {et~al.}(2013)\citenamefont {Dodson},
  \citenamefont {Hore},\ and\ \citenamefont {Wallace}}]{Hore1}%
  \BibitemOpen
  \bibfield  {author} {\bibinfo {author} {\bibfnamefont {C.~A.}\ \bibnamefont
  {Dodson}}, \bibinfo {author} {\bibfnamefont {P.~J.}\ \bibnamefont {Hore}},\
  and\ \bibinfo {author} {\bibfnamefont {M.~I.}\ \bibnamefont {Wallace}},\
  }\bibfield  {title} {\bibinfo {title} {A radical sense of direction:
  Signalling and mechachanism in cryptochrome magnetoreception},\ }\href@noop
  {} {\bibfield  {journal} {\bibinfo  {journal} {Trends in Biochemical
  Sciences}\ }\textbf {\bibinfo {volume} {38}} (\bibinfo {year}
  {2013})}\BibitemShut {NoStop}%
\bibitem [{\citenamefont {Wiltschko}\ and\ \citenamefont
  {Wiltschko}(2005)}]{inclination1}%
  \BibitemOpen
  \bibfield  {author} {\bibinfo {author} {\bibfnamefont {W.}~\bibnamefont
  {Wiltschko}}\ and\ \bibinfo {author} {\bibfnamefont {R.}~\bibnamefont
  {Wiltschko}},\ }\bibfield  {title} {\bibinfo {title} {Magnetic orientation
  and magnetoreception in birds and other animals},\ }\href@noop {} {\bibfield
  {journal} {\bibinfo  {journal} {Journal of Comparative Physiology A}\
  }\textbf {\bibinfo {volume} {191}},\ \bibinfo {pages} {675} (\bibinfo {year}
  {2005})}\BibitemShut {NoStop}%
\bibitem [{\citenamefont {Wiltschko}\ and\ \citenamefont
  {Wiltschko}(1972)}]{inclination2}%
  \BibitemOpen
  \bibfield  {author} {\bibinfo {author} {\bibfnamefont {W.}~\bibnamefont
  {Wiltschko}}\ and\ \bibinfo {author} {\bibfnamefont {R.}~\bibnamefont
  {Wiltschko}},\ }\bibfield  {title} {\bibinfo {title} {Magnetic compass of
  european robins},\ }\href@noop {} {\bibfield  {journal} {\bibinfo  {journal}
  {Science}\ }\textbf {\bibinfo {volume} {176}},\ \bibinfo {pages} {62}
  (\bibinfo {year} {1972})}\BibitemShut {NoStop}%
\bibitem [{\citenamefont {Wiltschko}\ and\ \citenamefont
  {Wiltschko}(2010)}]{inclination3}%
  \BibitemOpen
  \bibfield  {author} {\bibinfo {author} {\bibfnamefont {R.}~\bibnamefont
  {Wiltschko}}\ and\ \bibinfo {author} {\bibfnamefont {W.}~\bibnamefont
  {Wiltschko}},\ }\bibfield  {title} {\bibinfo {title} {Avian magnetic compass:
  its functional properties and physical basis},\ }\href@noop {} {\bibfield
  {journal} {\bibinfo  {journal} {Current Zoology}\ }\textbf {\bibinfo {volume}
  {56}},\ \bibinfo {pages} {265} (\bibinfo {year} {2010})}\BibitemShut
  {NoStop}%
\bibitem [{\citenamefont {Davila}\ \emph {et~al.}(2003)\citenamefont {Davila},
  \citenamefont {Fleissner}, \citenamefont {Winklhofer},\ and\ \citenamefont
  {Petersen}}]{inclination4}%
  \BibitemOpen
  \bibfield  {author} {\bibinfo {author} {\bibfnamefont {A.~F.}\ \bibnamefont
  {Davila}}, \bibinfo {author} {\bibfnamefont {G.}~\bibnamefont {Fleissner}},
  \bibinfo {author} {\bibfnamefont {M.}~\bibnamefont {Winklhofer}},\ and\
  \bibinfo {author} {\bibfnamefont {N.}~\bibnamefont {Petersen}},\ }\bibfield
  {title} {\bibinfo {title} {A new model for a magnetorecepter in homing
  pigeons based on interacting clusters of super-paramagnetic magnetite},\
  }\href@noop {} {\bibfield  {journal} {\bibinfo  {journal} {Physics and
  Chemistry of the Earth}\ }\textbf {\bibinfo {volume} {28}},\ \bibinfo {pages}
  {647} (\bibinfo {year} {2003})}\BibitemShut {NoStop}%
\bibitem [{\citenamefont {Wiltschko}\ \emph {et~al.}(1993)\citenamefont
  {Wiltschko}, \citenamefont {Munro}, \citenamefont {Ford},\ and\ \citenamefont
  {Wiltschko}}]{light1}%
  \BibitemOpen
  \bibfield  {author} {\bibinfo {author} {\bibfnamefont {W.}~\bibnamefont
  {Wiltschko}}, \bibinfo {author} {\bibfnamefont {U.}~\bibnamefont {Munro}},
  \bibinfo {author} {\bibfnamefont {H.}~\bibnamefont {Ford}},\ and\ \bibinfo
  {author} {\bibfnamefont {R.}~\bibnamefont {Wiltschko}},\ }\bibfield  {title}
  {\bibinfo {title} {Red light disrupts magnetic orientation of migratory
  birds},\ }\href@noop {} {\bibfield  {journal} {\bibinfo  {journal} {Nature}\
  }\textbf {\bibinfo {volume} {364}},\ \bibinfo {pages} {525} (\bibinfo {year}
  {1993})}\BibitemShut {NoStop}%
\bibitem [{\citenamefont {Wiltschko}\ and\ \citenamefont
  {Wiltschko}(1995)}]{light2}%
  \BibitemOpen
  \bibfield  {author} {\bibinfo {author} {\bibfnamefont {W.}~\bibnamefont
  {Wiltschko}}\ and\ \bibinfo {author} {\bibfnamefont {R.}~\bibnamefont
  {Wiltschko}},\ }\bibfield  {title} {\bibinfo {title} {Migratory orientation
  of european robins is affected by the wavelength of light as well as by a
  magnetic pulse},\ }\href@noop {} {\bibfield  {journal} {\bibinfo  {journal}
  {Journal of Comparative Physiology A}\ }\textbf {\bibinfo {volume} {177}},\
  \bibinfo {pages} {363} (\bibinfo {year} {1995})}\BibitemShut {NoStop}%
\bibitem [{\citenamefont {Wiltschko}\ and\ \citenamefont
  {Wiltschko}(1998)}]{light3}%
  \BibitemOpen
  \bibfield  {author} {\bibinfo {author} {\bibfnamefont {W.}~\bibnamefont
  {Wiltschko}}\ and\ \bibinfo {author} {\bibfnamefont {R.}~\bibnamefont
  {Wiltschko}},\ }\bibfield  {title} {\bibinfo {title} {Pigeon homing: Effect
  of various wavelengths of light during displacement},\ }\href@noop {}
  {\bibfield  {journal} {\bibinfo  {journal} {Naturwissenschaften}\ }\textbf
  {\bibinfo {volume} {85}},\ \bibinfo {pages} {164} (\bibinfo {year}
  {1998})}\BibitemShut {NoStop}%
\bibitem [{\citenamefont {Wiltschko}\ and\ \citenamefont
  {Wiltschko}(1999)}]{light4}%
  \BibitemOpen
  \bibfield  {author} {\bibinfo {author} {\bibfnamefont {W.}~\bibnamefont
  {Wiltschko}}\ and\ \bibinfo {author} {\bibfnamefont {R.}~\bibnamefont
  {Wiltschko}},\ }\bibfield  {title} {\bibinfo {title} {The effect of yellow
  and blue light on magnetic compass orientation in european robins,
  $erithacus$ $rubecula$},\ }\href@noop {} {\bibfield  {journal} {\bibinfo
  {journal} {Journal of Comparative Physiology A}\ }\textbf {\bibinfo {volume}
  {184}},\ \bibinfo {pages} {295} (\bibinfo {year} {1999})}\BibitemShut
  {NoStop}%
\bibitem [{\citenamefont {Wiltschko}\ \emph {et~al.}(2005)\citenamefont
  {Wiltschko}, \citenamefont {Ritz}, \citenamefont {Stapput}, \citenamefont
  {Thalau},\ and\ \citenamefont {Wiltschko}}]{light5}%
  \BibitemOpen
  \bibfield  {author} {\bibinfo {author} {\bibfnamefont {R.}~\bibnamefont
  {Wiltschko}}, \bibinfo {author} {\bibfnamefont {T.}~\bibnamefont {Ritz}},
  \bibinfo {author} {\bibfnamefont {K.}~\bibnamefont {Stapput}}, \bibinfo
  {author} {\bibfnamefont {P.}~\bibnamefont {Thalau}},\ and\ \bibinfo {author}
  {\bibfnamefont {W.}~\bibnamefont {Wiltschko}},\ }\bibfield  {title} {\bibinfo
  {title} {Two different types of light-dependent responses to magnetic fields
  in birds},\ }\href@noop {} {\bibfield  {journal} {\bibinfo  {journal}
  {Current Biology}\ }\textbf {\bibinfo {volume} {15}},\ \bibinfo {pages}
  {1518} (\bibinfo {year} {2005})}\BibitemShut {NoStop}%
\bibitem [{\citenamefont {Stapput}\ \emph {et~al.}(2008)\citenamefont
  {Stapput}, \citenamefont {Thalau}, \citenamefont {Wiltschko},\ and\
  \citenamefont {Wiltschko}}]{light6}%
  \BibitemOpen
  \bibfield  {author} {\bibinfo {author} {\bibfnamefont {K.}~\bibnamefont
  {Stapput}}, \bibinfo {author} {\bibfnamefont {P.}~\bibnamefont {Thalau}},
  \bibinfo {author} {\bibfnamefont {R.}~\bibnamefont {Wiltschko}},\ and\
  \bibinfo {author} {\bibfnamefont {W.}~\bibnamefont {Wiltschko}},\ }\bibfield
  {title} {\bibinfo {title} {Orientation of birds in total darkness},\
  }\href@noop {} {\bibfield  {journal} {\bibinfo  {journal} {Current Biology}\
  }\textbf {\bibinfo {volume} {18}},\ \bibinfo {pages} {602} (\bibinfo {year}
  {2008})}\BibitemShut {NoStop}%
\bibitem [{\citenamefont {Wiltschko}\ \emph {et~al.}(2011)\citenamefont
  {Wiltschko}, \citenamefont {Denzau}, \citenamefont {Gehring}, \citenamefont
  {Thalau},\ and\ \citenamefont {Wiltschko}}]{light7}%
  \BibitemOpen
  \bibfield  {author} {\bibinfo {author} {\bibfnamefont {R.}~\bibnamefont
  {Wiltschko}}, \bibinfo {author} {\bibfnamefont {S.}~\bibnamefont {Denzau}},
  \bibinfo {author} {\bibfnamefont {D.}~\bibnamefont {Gehring}}, \bibinfo
  {author} {\bibfnamefont {P.}~\bibnamefont {Thalau}},\ and\ \bibinfo {author}
  {\bibfnamefont {W.}~\bibnamefont {Wiltschko}},\ }\bibfield  {title} {\bibinfo
  {title} {Magnetic orientation of migratory robins, $erithacus$ $rubecula$,
  under long-wavelength light},\ }\href@noop {} {\bibfield  {journal} {\bibinfo
   {journal} {Journal of Experimental Biology}\ }\textbf {\bibinfo {volume}
  {214}},\ \bibinfo {pages} {3096} (\bibinfo {year} {2011})}\BibitemShut
  {NoStop}%
\bibitem [{\citenamefont {Wiltschko}\ and\ \citenamefont
  {Wiltschko}(1978)}]{range1}%
  \BibitemOpen
  \bibfield  {author} {\bibinfo {author} {\bibfnamefont {R.}~\bibnamefont
  {Wiltschko}}\ and\ \bibinfo {author} {\bibfnamefont {W.}~\bibnamefont
  {Wiltschko}},\ }\bibfield  {title} {\bibinfo {title} {Evidence for the use of
  magnetic outward-journey information in homing pigeons},\ }\href@noop {}
  {\bibfield  {journal} {\bibinfo  {journal} {Naturwissenschaften}\ }\textbf
  {\bibinfo {volume} {65}},\ \bibinfo {pages} {112} (\bibinfo {year}
  {1978})}\BibitemShut {NoStop}%
\bibitem [{\citenamefont {Gauger}\ \emph {et~al.}(2011)\citenamefont {Gauger},
  \citenamefont {Rieper}, \citenamefont {Morton}, \citenamefont {Benjamin},\
  and\ \citenamefont {Vedral}}]{Gauger1}%
  \BibitemOpen
  \bibfield  {author} {\bibinfo {author} {\bibfnamefont {E.~M.}\ \bibnamefont
  {Gauger}}, \bibinfo {author} {\bibfnamefont {E.}~\bibnamefont {Rieper}},
  \bibinfo {author} {\bibfnamefont {J.~J.~L.}\ \bibnamefont {Morton}}, \bibinfo
  {author} {\bibfnamefont {S.~C.}\ \bibnamefont {Benjamin}},\ and\ \bibinfo
  {author} {\bibfnamefont {V.}~\bibnamefont {Vedral}},\ }\bibfield  {title}
  {\bibinfo {title} {Sustained quantum coherence and entanglement in the avian
  compass},\ }\href {https://doi.org/10.1103/PhysRevLett.106.040503} {\bibfield
   {journal} {\bibinfo  {journal} {Phys. Rev. Lett.}\ }\textbf {\bibinfo
  {volume} {106}},\ \bibinfo {pages} {040503} (\bibinfo {year}
  {2011})}\BibitemShut {NoStop}%
\bibitem [{\citenamefont {Levy}\ and\ \citenamefont
  {Shalit}(2014)}]{levy2014dilation}%
  \BibitemOpen
  \bibfield  {author} {\bibinfo {author} {\bibfnamefont {E.}~\bibnamefont
  {Levy}}\ and\ \bibinfo {author} {\bibfnamefont {O.~M.}\ \bibnamefont
  {Shalit}},\ }\bibfield  {title} {\bibinfo {title} {Dilation theory in finite
  dimensions: the possible, the impossible and the unknown},\ }\href@noop {}
  {\bibfield  {journal} {\bibinfo  {journal} {Rocky Mountain Journal of
  Mathematics}\ }\textbf {\bibinfo {volume} {44}},\ \bibinfo {pages} {203}
  (\bibinfo {year} {2014})}\BibitemShut {NoStop}%
\bibitem [{\citenamefont {Kominis}(2015)}]{kominis2015radical}%
  \BibitemOpen
  \bibfield  {author} {\bibinfo {author} {\bibfnamefont {I.~K.}\ \bibnamefont
  {Kominis}},\ }\bibfield  {title} {\bibinfo {title} {The radical-pair
  mechanism as a paradigm for the emerging science of quantum biology},\
  }\href@noop {} {\bibfield  {journal} {\bibinfo  {journal} {Modern Physics
  Letters B}\ }\textbf {\bibinfo {volume} {29}},\ \bibinfo {pages} {1530013}
  (\bibinfo {year} {2015})}\BibitemShut {NoStop}%
\bibitem [{\citenamefont {Zadeh-Haghighi}\ and\ \citenamefont
  {Simon}(2022)}]{zadeh2022magnetic}%
  \BibitemOpen
  \bibfield  {author} {\bibinfo {author} {\bibfnamefont {H.}~\bibnamefont
  {Zadeh-Haghighi}}\ and\ \bibinfo {author} {\bibfnamefont {C.}~\bibnamefont
  {Simon}},\ }\bibfield  {title} {\bibinfo {title} {Magnetic field effects in
  biology from the perspective of the radical pair mechanism},\ }\href@noop {}
  {\bibfield  {journal} {\bibinfo  {journal} {Journal of the Royal Society
  Interface}\ }\textbf {\bibinfo {volume} {19}},\ \bibinfo {pages} {20220325}
  (\bibinfo {year} {2022})}\BibitemShut {NoStop}%
\end{thebibliography}%
\clearpage

\appendix
\newpage

\section{Quantum Simulation Details}
\label{qs_details}

Here we give an example of the quantum circuit of the $M_1 = \sqrt{k_d\delta t}P_1$ at the first time step, $5 \times 10^{-5}s$ to evolve the states. After multiplied by the unitary matrix accounting for the coherent part as in Eq. (\ref{eq:me}), $M_1$ becomes a $10\times10$ sparse matrix shown in Eq. (\ref{m1_matrix}) and the non-zero values are shown in Eq. (\ref{m1_values}).  

\begin{equation}
\begin{bmatrix}
0  & 0 & 0 & 0 & 0 & 0 & 0 & 0 & 0 & 0 \\
0  & 0 & 0 & 0 & 0 & 0 & 0 & 0 & 0 & 0 \\
0  & 0 & 0 & 0 & 0 & 0 & 0 & 0 & 0 & 0 \\
0  & 0 & 0 & 0 & 0 & 0 & 0 & 0 & 0 & 0 \\
0  & 0 & 0 & 0 & 0 & 0 & 0 & 0 & 0 & 0 \\
0  & 0 & 0 & 0 & 0 & 0 & 0 & 0 & 0 & 0 \\
0  & 0 & 0 & 0 & 0 & 0 & 0 & 0 & 0 & 0 \\
0  & 0 & 0 & 0 & 0 & 0 & 0 & 0 & 0 & 0 \\
m_1 & m_2 & m_3 & m_4 & m_5 & m_6 & m_7 & m_8 & 0 & 0 \\
0  & 0 & 0 & 0 & 0 & 0 & 0 & 0 & 0 & 0 \\
\end{bmatrix}
\label{m1_matrix}
\end{equation}

\begin{equation}
\begin{aligned}
  m_1 = -5.22\times10^{-2} - 4.05\times10^{-1}i \\
  m_2 = -6.53\times10^{-4} + 1.23\times10^{-2}i \\
  m_3 = -2.93\times10^{-1} + 3.80\times10^{-3}i \\
  m_4 = -1.70\times10^{-2} + 1.38\times10^{-4}i \\
  m_5 = 2.93\times10^{-1} + 3.78\times10^{-3}i \\
  m_6 = -7.39\times10^{-1} -2.62\times10^{-3}i \\
  m_7 = -5.22\times10^{-3} + 4.05\times10^{-1}i \\
  m_8 = 5.63\times10^{-4}+5.37\times10^{-3}i \\
  \label{m1_values}
\end{aligned}
\end{equation}

\begin{figure}
    \centering
        \includegraphics[width=0.4\textwidth, height=20cm]{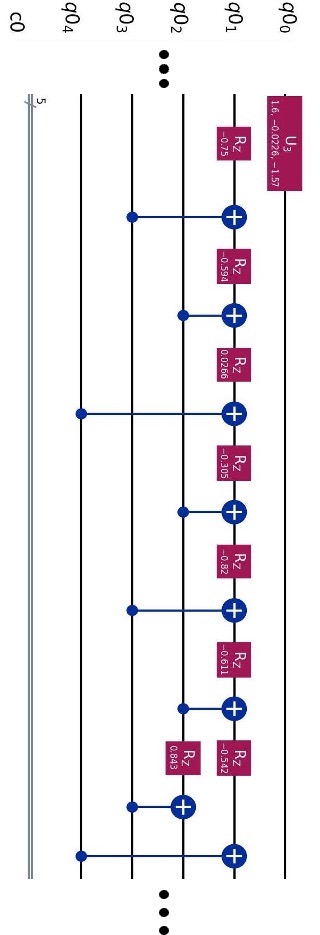}
        \caption{An example of a portion of the quantum gate sequence of a Kraus operator. This is only a small portion of the circuit. The full circuit has 2097 gates and the details will be available from the authors on reasonable request.}
    \label{fig:circuit}
\end{figure}

After we apply the unitary dilation described in Eq. (\ref{eq:1-dialtion}) on the $M_1$, we obtain a $20\times20$ unitary matrix $U_{M_1}$. To simulate this operation on quantum simulator, we have to use 5 qubits to cover the 20 dimensions. Leveraging the Qiskit's transpile function (qiskit.compliler.transpile), we decomposed the resulted unitary operator $U_{M_1}$ to 2097 gates where a portion of the circuit is shown below (Fig. \ref{fig:circuit}) using the basis gates: `u3', `cx' and `rz' as shown in Eq.\eqref{eq:basis gates} , on the backend of qasm\_simulator. The details of the decomposition of all the quantum circuits used are available on reasonable request.
\begin{align}\label{eq:basis gates}
    \begin{aligned}\begin{split}U_3(\theta, \phi, \lambda) =
    \begin{pmatrix}
        \cos\left(\frac{\theta}{2}\right)          & -e^{i\lambda}\sin\left(\frac{\theta}{2}\right) \\
        e^{i\phi}\sin\left(\frac{\theta}{2}\right) & e^{i(\phi+\lambda)}\cos\left(\frac{\theta}{2}\right)
    \end{pmatrix}\end{split}\\
     \begin{split}CX&=
    I \otimes |0\rangle\langle0| + X \otimes |1\rangle\langle1| =
    \begin{pmatrix}
        1 & 0 & 0 & 0 \\
        0 & 0 & 0 & 1 \\
        0 & 0 & 1 & 0 \\
        0 & 1 & 0 & 0
    \end{pmatrix}\end{split}\\
    \begin{split}
        Rz(\theta) = 
        \begin{pmatrix}
            e^{-i\frac{\theta}{2}} & 0 \\
            0 & e^{i\frac{\theta}{2}}
        \end{pmatrix}
    \end{split}
    \end{aligned}
\end{align}

\clearpage


\end{document}